\documentclass[aps,prl,twocolumn,groupedaddress,showpacs]{revtex4}

\usepackage{graphicx,color}
\usepackage{amsmath}

\begin{document}
\title{The nature of highly anisotropic free-electron-like states
in a glycinate monolayer on Cu(100)}

\author{Matthew S. Dyer and Mats Persson}
\address{The Surface Science Research Centre, The University of Liverpool, Liverpool, L69 3BX, UK}
\title{The nature of highly anisotropic free-electron-like states
in a glycinate monolayer on Cu(100)}

\begin{abstract}
The free-electron-like state observed in a scanning tunneling  spectroscopy study of a chiral p(2$\times$4) monolayer of glycinate ions  on the Cu(100) surface [K. Kanazawa {\itshape et al},  {\itshape J. Am. Chem. Soc.} {\bfseries 2007}, {\itshape 129}, 740] is shown from density functional theory calculations to originate from a Cu Shockley surface state at the surface Brillouin zone boundary of the clean surface with highly anisotropic dispersion.  The presence of the glycinate ions on the surface causes a dramatically enhanced tunneling into this surface state that is otherwise not observed in tunneling on the bare surface.
\end{abstract}

\maketitle

Recently, free-electron-like states arising in layers of organic molecules on metal surfaces at energies close to the Fermi energy have been observed by scanning tunneling and photoemission spectroscopies~\cite{Nicetal06,temirovn06,kanazawajacs07}. Interest in such delocalized electronic states is high, because of their potential use in molecular electronics applications. In systems where the molecule is more or less physisorbed the observed free-electron like state was simply identified as a metal surface state~\cite{Nicetal06}, whereas in systems where the molecule is chemisorbed the origin of the state is not properly understood~\cite{temirovn06,kanazawajacs07}.

In the scanning tunneling spectroscopy (STS) study by Kanazawa  {\itshape et al}\cite{kanazawajacs07} an anisotropic free-electron-like state was observed in a p(2$\times$4) monolayer of glycinate anions on the Cu(100) surface just above the Fermi energy. From an analysis of the two-dimensional autocorrelations of the differential conductance ($\frac{dI}{dV}$) images , they were able to show that the level of dispersion in the [110] and [\=110] directions differs by an order of magnitude. From the absence of any observation of a surface state on the clean surface in the $\frac{dI}{dV}$ spectra, they suggested that the state originated from an interaction between the glycinate monolayer and the Cu surface and its anisotropic dispersion from the different strength of interaction between neighboring molecules in the two directions. There is need to carry out electronic structure calculations to clarify the origin of this state.

In this letter we present the results of density functional theory (DFT) calculations on the p(2$\times$4) glycinate monolayer and show that the origin of the anisotropic free-electron-like state (AFES) is a Cu Shockley surface state (SS). Although Kanazawa {\itshape et al} did not observe a SS near the Fermi energy on the bare Cu (100) surface~\cite{kanazawajacs07} a SS has been identified near the Fermi energy at the surface Brillouin zone (SBZ) boundary~\cite{kevanprb83,eucedaprb83,baldacchiniprb03,sekibaprb07}. We show that the SS is folded back to the $\bar\Gamma$ point in the (SBZ) for the p(2$\times$4) structure studied by Kanazawa {\itshape et al}, and that the anisotropic parabolic dispersion of the observed AFES  is due to enhanced tunneling into the SS, mediated by the lowest unoccupied molecular orbitals of the glycinate ions.

\begin{figure}[h]
  (a)\\\includegraphics[height=3cm]{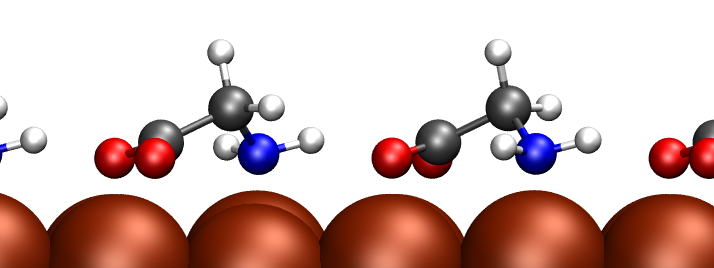}\\

  \begin{minipage}[t]{4cm}(b)\\

    \includegraphics[height=4cm]{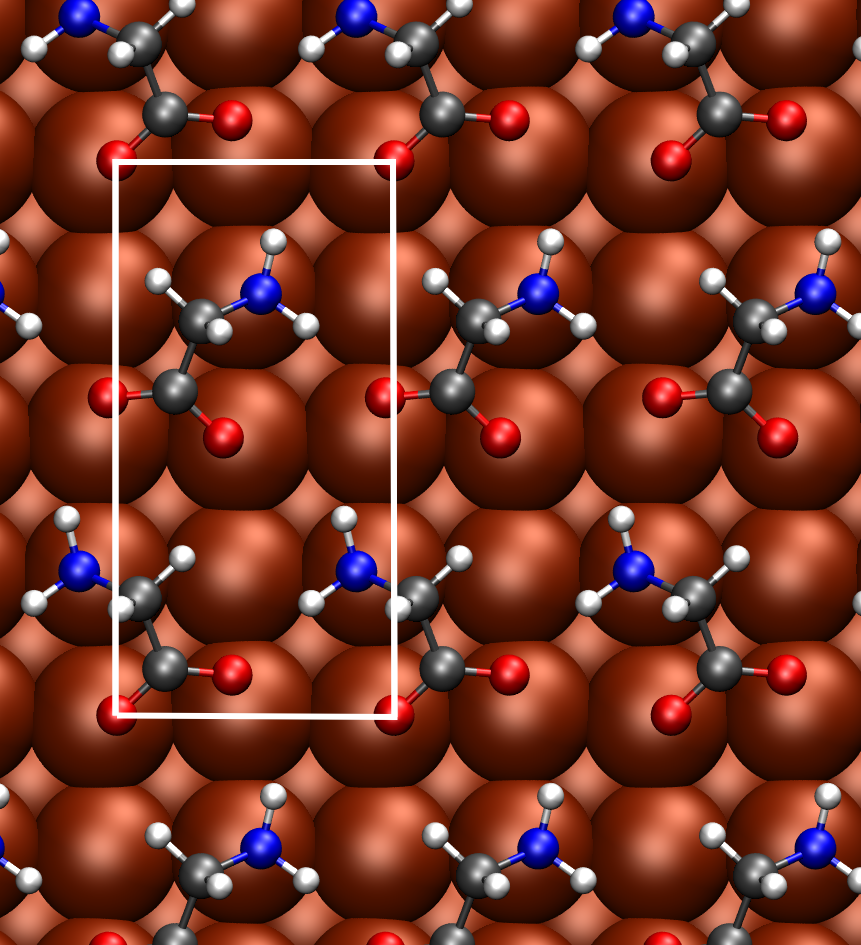}\\
  \end{minipage}
  \begin{minipage}[t]{4cm}(c)\\

    \begin{center}
      \setlength{\unitlength}{0.33333cm}
      \begin{picture}(12,12)
         \put(2,2){\framebox(8,8){}}
         \put(6,6){\vector(1,0){4}}
         \put(4,4){\makebox(2,2)[tr]{$\bar\Gamma$}}
         \put(5,10){\makebox(2,2)[l]{\=Y$'$}}
         \put(10,5){\makebox(2,2)[l]{\=X$'$}}
         \put(10,10){\makebox(2,2)[lb]{\=M}}
         \put(4,5){\framebox(4,2){}}
         \put(6,6){\vector(1,0){2}}
         \put(6,6){\vector(0,1){1}}
         \put(6,6){\vector(0,1){4}}
         \put(8,6){\makebox(2,2)[lb]{\=X}}
         \put(4,7){\makebox(2,2)[rb]{\=Y}}
      \end{picture}
    \end{center}
  \end{minipage}
  \caption{Calculated adsorption geometry of glycinate molecules on Cu (100) (a) side view and  (b) top view. The 2$\times$4 unit cell is shown from above in white. (c) Labels for high symmetry points in the square SBZ of the Cu(100) surface and the rectangular SBZ of the p(2$\times$4) glycinate system. The $\bar\Gamma$-\=X direction corresponds to the direction in real space labeled [110] by Kanazawa {\itshape et al}\protect\cite{kanazawajacs07}, and $\bar\Gamma$-\=Y the [\=110] direction. The \=X$'$ and \=Y$'$ points in the substrate SBZ fold back to the $\bar\Gamma$ point in the monolayer SBZ. (color online)
   \label{figgeom1}\label{figSBZ1}}
\end{figure}
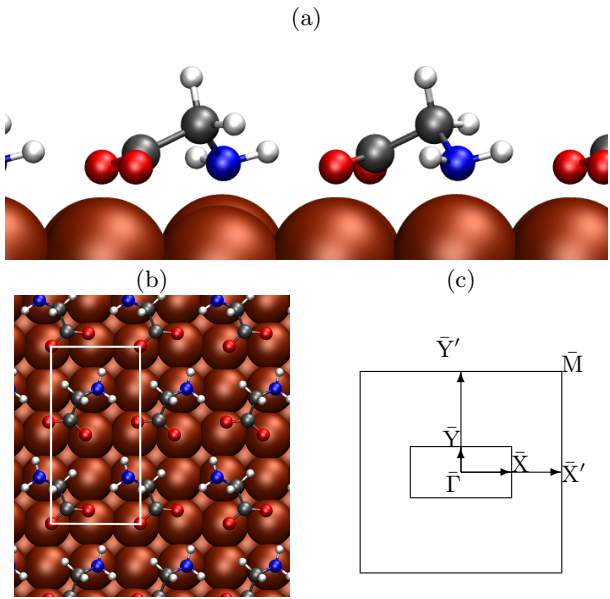

Glycine ($\mathrm{NH_2CH_2COOH}$) is the simplest amino acid and its adsorption geometry on the Cu (100) surface is well understood following several studies in the last ten years with scanning tunneling microscopy (STM)~\cite{kanazawajacs07,zhaoss99,zhaomsec01}, photoelectron diffraction~\cite{kangjcp03} and DFT calculations~\cite{maess04}. The glycine molecule loses a hydrogen atom on adsorption becoming a glycinate anion ($\mathrm{NH_2CH_2COO^-}$) which then binds to the Cu surface through the nitrogen atom and two oxygen atoms in a tridentate fashion. The p(2$\times$4) structure~\cite{kanazawajacs07,zhaomsec01} has alternating rows of the left and right enantiomers of glycinate ions, the rows propagating along the close-packed rows of Cu atoms (see Fig.~\ref{figgeom1}). In addition to the hetereochiral p(2$\times$4) structure a homochiral c(2$\times$4)~\cite{zhaoss99,zhaomsec01,kanazawaprl07} structure is also formed at the same temperature on the Cu(100) surface.

Density functional theory (DFT) calculations were carried out using
the plane-wave based VASP code~\cite{kresseprb96}. Calculations were made using the projector augmented wave method~\cite{kresseprb99} and the exchange--correlation functional using the generalized gradient approximation proposed by Perdew and Wang~\cite{perdewprb92}. Glycinate ions were placed on a six layer Cu(100) slab in a 5.14$\times$10.28$\times$25 \AA$^3$ supercell. A full geometry relaxation was carried out with a plane wave cutoff of 400 eV on a 8$\times$4$\times$1 $k$-point grid until the forces acting on the ions were smaller in magnitude than 0.01 eV/{\AA}. The atoms in the top three Cu layers were also allowed to relax. Calculations were also performed with a 12 layer bare and adsorbate-covered Cu slab to ensure that effects caused by the finite thickness of the slab were not important. 

Following the commonly used Tersoff--Hamann approximation~\cite{tersoffprl83} the differential conductance measured in STS experiments was approximated using the local density of states (LDOS) calculated at the position of the tip, which was set at 7 {\AA} from the top layer of Cu atoms. Once the relaxed geometry had been found, the local density of states (LDOS) and molecular orbital projected density of states (MO-PDOS) was calculated on the original $k$-point grid and at 20 points in lines from the $\bar\Gamma$ point to both the \=X and \=Y points in the SBZ (see Fig. \ref{figSBZ1}(c)). These calculations used a charge density calculated self-consistently on a 16$\times$8$\times$1 $k$-point grid. The MO-PDOS was calculated by projecting the density of states of the full system onto the wavefunctions for the isolated monolayer in the same supercell at each $k$-point.

Fig.~\ref{figgeom1} shows the adsorption geometry of the glycinate molecules following geometry relaxation from one close to that found by Mae and Morikawa\cite{maess04} for the p(2$\times$4) structure. The resulting Cu-N bond length of 2.10 {\AA} and the Cu-O bond lengths of 2.08 {\AA} and 2.18 {\AA} are in good agreement with previous theoretical\cite{maess04} and experimental\cite{kangjcp03} findings.

The electronic structure of molecular monolayer was investigated by calculating the MO-PDOS of the adsorbed system along with the band decomposed DOS of an isolated layer of neutral glycinate radicals in the same geometry as on the Cu surface. The isolated layer is found to have a partially occupied band with a width of approximately 0.4 eV across the SBZ formed by the overlap of 
molecular orbitals on neighboring molecules. Projection of the DOS of the full adsorbed system onto the partially occupied band in the isolated glycinate layer shows that the band becomes fully occupied in the presence of the Cu surface forming negatively 
charged glycinate anions in agreement with previous 
studies~\cite{kanazawajacs07,zhaomsec01,kangjcp03,barlowss98}.
The glycinate states are considerably broadened upon adsorption reflecting a strong interaction between the glycinate anions and the Cu surface. Although bands are formed due to interactions between neighboring molecules they lie too low in energy and do not have the correct dispersion behavior to explain free-electron-like state observed by Kanazawa {\itshape et al}~\cite{kanazawajacs07}.

\begin{figure}[h]
  \begin{minipage}[t]{4cm}(a)\\
     \includegraphics[height=8.89cm]{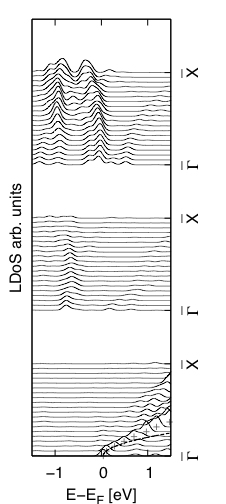}
  \end{minipage}
  \begin{minipage}[t]{4cm}(b)\\
     \includegraphics[height=8.89cm]{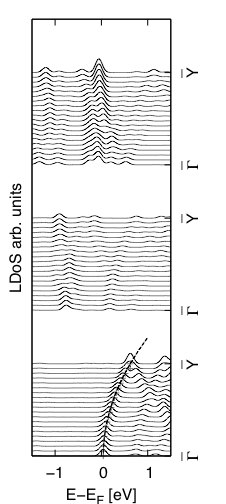}
  \end{minipage}
  \caption{K-point decomposed LDOS at 7 {\AA} above the Cu surface in the (a) $\bar\Gamma$-\=X direction and the (b) $\bar\Gamma$-\=Y direction.  The dashed curves are parabolic dispersions with the effective masses observed by Kanazawa {\itshape et al}~\cite{kanazawajacs07}. The lower set of curves are for the fully relaxed adsorbate geometry, the middle set with the glycinate ions displaced 0.5 {\AA} away from the Cu surface and the upper set 1.0 {\AA} away. Data from neighboring $k$-points have been displaced vertically   and crosses mark the position of the top of the peak at its base. 
  \label{figLDOSdisp1}\label{figLDOSdisp2}}
\end{figure}

The calculated LDOS shows a peak which has an anisotropic 
free-electron-like dispersion in very good agreement to that found by Kanazawa {\itshape et al}~\cite{kanazawajacs07} (Figure~\ref{figLDOSdisp1}). They found that the dispersion of the AFES giving rise to the peak in the auto-correlated $\frac{dI}{dV}$ spectra was well described by parabolic dispersion with effective masses of 0.061 $\mathrm{m_e}$ and 0.61 $\mathrm{m_e}$ in the [110] and the [\=110] directions, respectively. The dispersion of the peak in the calculated LDOS shown in Fig.~\ref{figLDOSdisp1}  agrees very well with the experimentally observed parabolic dispersion, although the experimentally observed onset of the peak was at 130 meV above the Fermi energy slightly higher than the calculated value of 40 meV. The origin of the AFES behind the calculated peak can be understood from 
a scrutiny of the electronic states of the bare Cu slab.

\begin{figure}[h]
\includegraphics[height=5cm]{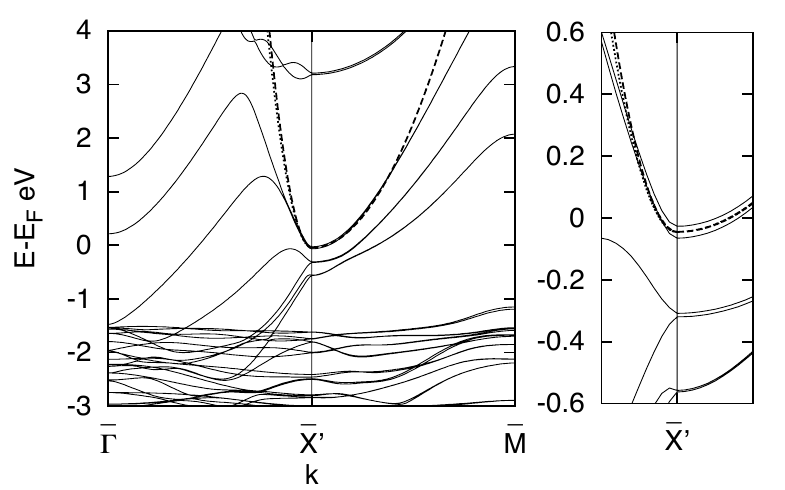}
\caption{Bandstructure of a six layer Cu(100) slab. The dotted curve shows a parabolic dispersion with the effective mass of the surface state on the bare Cu surface measured by Kevan\cite{kevanprb83} and the dashed curves with the effective masses observed by Kanazawa {\itshape et al} for the state with the glycinate monolayer present~\protect\cite{kanazawajacs07}.
\label{figCubands1}}
\end{figure}

Calculations show that the SS of the bare Cu slab has a similar energy and anisotropic dispersion to the AFES. The SS is clearly observable at the \=X$'$ point in the calculated band structure shown in Fig.~\ref{figCubands1}. Two degenerate SS are found on each surface of the bare slab; one at the \=X$'$ point and one at the equivalent \=Y$'$ point for the bare surface.  The energy splitting in the SS at the \=X$'$ point (see Table~\ref{tabenergies1}) due to interactions between the two different sides of the slab is very small even with the six layer slab showing that the SS decays quickly into the bulk. The calculated dispersion of the SS near the \=X$'$ point reproduces closely the observed parabolic dispersion with an effective mass of 0.067$\pm$0.01 $\mathrm{m_e}$ in the [110] direction measured by Kevan\cite{kevanprb83} in an angle resolved photoemission study. The calculated SS energy of 45 meV below the Fermi energy at the \=X$'$ point is also in good agreement with the measured onset of 58$\pm$0.05 meV below the Fermi energy~\cite{kevanprb83}.  The effective mass of 0.067$\pm$0.01 $\mathrm{m_e}$ in the [110] direction of the SS is close the measured effective mass of 0.061 $\mathrm{m_e}$ of the AFES in the same direction. Furthermore, the calculated SS in Fig.~\ref{figCubands1} shows also a parabolic-like dispersion in the [1\=10] direction in close agreement with the measured dispersion of the AFES in Fig.~\ref{figCubands1} in the same surface direction with an effective mass of 0.61 $\mathrm{m_e}$.

The close agreement of the calculated dispersions of the SS and the AFES suggests that the observed AFES is due to tunneling into the SS. Tunneling into the SS must be enhanced by the monolayer, however, since the SS cannot be observed on the bare surface using STS~\cite{kanazawajacs07}. Kanazawa {\itshape et al} observed a peak in the $\frac{dI}{dV}$ spectrum over the glycinate monolayer with the steep onset and gradual decay with increasing bias characteristic of a dispersive state, but no such peak was observed in the $\frac{dI}{dV}$ spectrum over the clean Cu surface. Calculations of the LDOS at 7 {\AA} above the bare Cu slab show no contribution due to the SS, only contributions from bulk states around the $\Gamma$-point. On the bare substrate the SS at the \=X$'$ and \=Y$'$ points decay too rapidly into the vacuum region to be observed. 

That the calculated AFES arises solely from the molecular states in the monolayer is ruled out by the behavior of the calculated LDOS when changing the distance of the glycinate ion monolayer from the Cu surface by 0.5 {\AA} and 1 {\AA} without subsequent relaxation of the geometry. The peak with anisotropic dispersion in LDOS  (Figure \ref{figLDOSdisp2}) is seen to get gradually smaller as the glycinate ions are withdrawn from the surface. If the peak were only due to tunneling into the Cu surface then the height of the peak should stay constant.  A peak due to tunneling through a purely molecular state would get taller as the molecules were brought closer to the tip. This behavior can be seen in the peaks below the Fermi energy. The peak with anisotropic dispersion gets smaller as the molecules are withdrawn showing that AFES is neither a purely molecular state or SS. However, this result suggests that enhanced tunneling into the SS arises from an interaction between the glycinate ions and the Cu surface. 

An interaction of the SS with the glycinate monolayer is confirmed  by the fact that the degeneracy of the SS at the \=X$'$ and the \=Y$'$ points has been lifted and their energies are shifted up by 85 meV and 148 meV, respectively, upon adsorption. The SS arising from both the \=X$'$ and \=Y$'$ points on the bare surface are folded back to the $\bar\Gamma$ point in the 2$\times$4 unit cell. They can be distinguished by comparing their dispersion in the $\bar\Gamma$-\=X ($\parallel$[110]) and $\bar\Gamma$-\=Y ($\parallel$[\=110]) directions. The SS arising from the \=X$'$ point has a smaller effective mass in the $\bar\Gamma$-\=X direction and that arising from the \=Y$'$ has a smaller effective mass in the $\bar\Gamma$-\=Y direction. The origin of these adsorption-induced energy shifts of the SS was investigated by calculating the SS energy of the frozen bare slab where Cu atoms were held in the geometry they had in the presence the glycinate anions. As shown in Table \ref{tabenergies1}, these energies show that although part of the increase in energy of the Shockley surface state is due to the adsorbate-induced changes in geometry of the Cu surface, part of the energy shift is only due to the presence of the glycinate anions. A substantial part of the shift in energy must therefore be due to direct electronic interaction between the SS and the glycinate anions states. The energies of the SS  from both the \=X$'$ and \=Y$'$ points on the bottom side of the slab remains roughly constant and are degenerate, showing that the slab is thick enough to prevent interactions between the monolayer and the SS at the bottom of the slab.

\begin{table} 
\caption{Calculated energies of the Shockley surface state. 
\label{tabenergies1}}
\begin{ruledtabular}
\begin{tabular}{l|c|c|c}
        & Relaxed & Frozen & Glycinate \\ 
Surface, k-point & bare slab & bare slab\footnote{Six layer
             Cu slab with the Cu atoms held in the geometry they 
             have when the glycinate atoms are present} & monolayer \\ \hline
       Top, \=X$'$  & -45\footnote{All energies in meV relative to the Fermi energy of the system.} & +3 & +40 \\
       Top, \=Y$'$  & -45 & -15 & +103 \\
       Bottom, \=X$'$  & -8 & -5 & -5 \\
       Bottom, \=Y$'$  & -8 & -5 & -5
    \end{tabular}
  \end{ruledtabular}
\end{table}

A most interesting result is that only the SS arising from the \=X$'$ point is responsible for the AFES in the LDOS in spite of the fact that a larger energy shift is found for the SS arising from the \=Y$'$ point, indicating a stronger interaction with the glycinate anions. This result explains nicely the observation in the STS experiments~\cite{kanazawajacs07} that the AFES has a smaller effective mass in the $\bar\Gamma$-\=X direction than in the $\bar\Gamma$-\=Y direction. The SS arising from the \=Y$'$ point would have a higher effective mass in $\bar\Gamma$-\=X direction than in the $\bar\Gamma$-\=Y direction. That only the SS from the \=X$'$ point contribute to the LDOS is consistent with the suggestion that the LUMO of the glycinate 
anions plays an important r\^ole in the enhanced tunneling into the SS. The separate  contributions to the MO-PDOS from different $k$-points in the  $\bar\Gamma$-\=X and $\bar\Gamma$-\=Y directions show that the SS giving rise to the AFES in the LDOS has non-zero overlap with several of the glycinate anion states. The largest overlap is with the first fully unoccupied band in the isolated monolayer, the LUMO states of the glycinate anions. In contrast, the SS arising from the \=Y$'$ point has no discernible overlap with these molecular orbitals. This lack of overlap is not simply a symmetry effect. The monolayer has only one non-trivial symmetry operation: a combined translation by two Cu lattice spacings in the $\bar\Gamma$-\=Y and a reflection. Because of the predominant $p$ character of the SS at the  \=X$'$ and \=Y$'$ points, they are both even under this symmetry operation. 

\begin{figure}[ht]
\includegraphics[height=5cm]{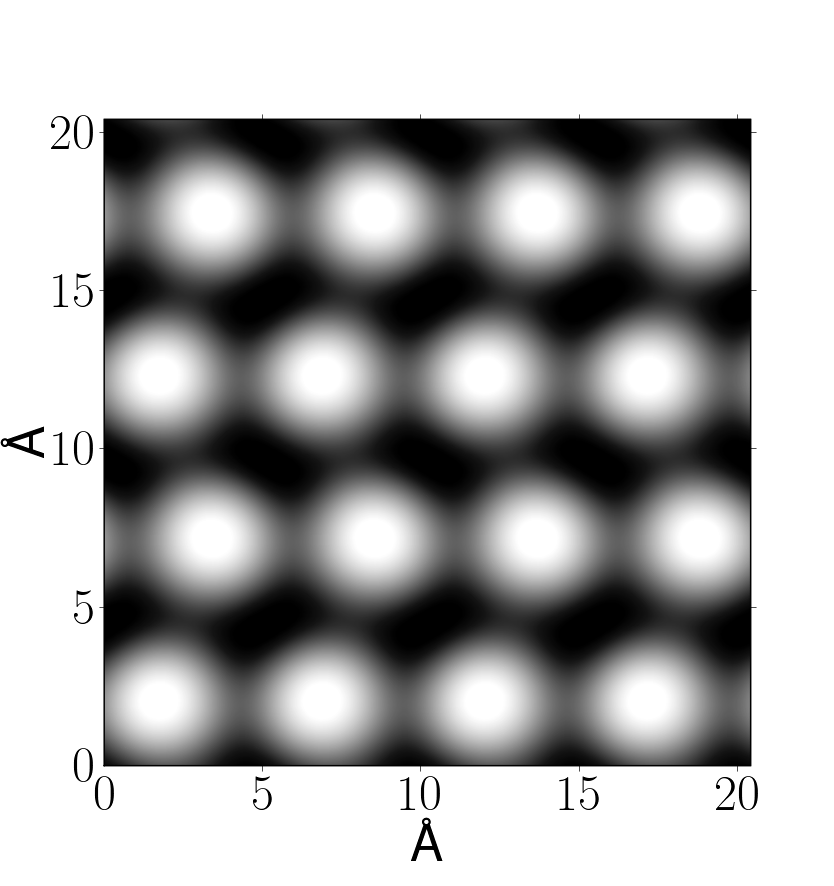}
\caption{Topography of constant orbital density of the AFES at the $\Gamma$ point with an average distance of 7.0 {\AA} from the Cu surface. Black (white) represents a height of 6.9 (7.2) {\AA} from the Cu surface. Each peak is centered over single glycinate anion.\label{figgammaldos}}
\end{figure}

The importance of the glycinate monolayer for the enhanced tunneling into the SS is illustrated by the plot in Fig.~\ref{figgammaldos} of the topography of the orbital density of the AFES at
the $\Gamma$-point in a plane located 7{\AA} away from the surface. This plot shows that the tunneling into the SS is localized to the glycinate ions and  centered over the protruding hydrogen atom. The localization of the tunneling above the glycinate ions and the overlap of only the SS  arising from the with \=X$'$ point with the LUMO suggests that this small overlap is sufficient for resonant tunneling through the LUMO into the SS beneath the molecules. 

In conclusion,  we have shown from our density functional calculations that an organic molecular monolayer can dramatically enhance the tunneling into a metal surface state that is otherwise not observed in scanning tunneling spectroscopy (STS). This finding is of importance in understanding the origin of free-electron like states in tunneling through molecular monolayers.

\begin{acknowledgments}
The authors would like to thank the Marie Curie Research Training Network PRAIRIES, contract MRTN-CT-2006-035810, and the Swedish Research Council (VR) for financial support and MSD for funding of postdoctoral fellowship and computational resources by the the University of Liverpool.
\end{acknowledgments}

\bibliography{liverpool}

\begin{thebibliography}{17}
\expandafter\ifx\csname natexlab\endcsname\relax\def\natexlab#1{#1}\fi
\expandafter\ifx\csname bibnamefont\endcsname\relax
  \def\bibnamefont#1{#1}\fi
\expandafter\ifx\csname bibfnamefont\endcsname\relax
  \def\bibfnamefont#1{#1}\fi
\expandafter\ifx\csname citenamefont\endcsname\relax
  \def\citenamefont#1{#1}\fi
\expandafter\ifx\csname url\endcsname\relax
  \def\url#1{\texttt{#1}}\fi
\expandafter\ifx\csname urlprefix\endcsname\relax\def\urlprefix{URL }\fi
\providecommand{\bibinfo}[2]{#2}
\providecommand{\eprint}[2][]{\url{#2}}

\bibitem[{\citenamefont{Kanazawa
  et~al.}(2007{\natexlab{a}})\citenamefont{Kanazawa, Sainoo, Konishi, Yoshida,
  Taninaka, Okada, Berthe, Kobayashi, Takeuchi, and
  Shigekawa}}]{kanazawajacs07}
\bibinfo{author}{\bibfnamefont{K.}~\bibnamefont{Kanazawa}},
  \bibinfo{author}{\bibfnamefont{Y.}~\bibnamefont{Sainoo}},
  \bibinfo{author}{\bibfnamefont{Y.}~\bibnamefont{Konishi}},
  \bibinfo{author}{\bibfnamefont{S.}~\bibnamefont{Yoshida}},
  \bibinfo{author}{\bibfnamefont{A.}~\bibnamefont{Taninaka}},
  \bibinfo{author}{\bibfnamefont{A.}~\bibnamefont{Okada}},
  \bibinfo{author}{\bibfnamefont{M.}~\bibnamefont{Berthe}},
  \bibinfo{author}{\bibfnamefont{N.}~\bibnamefont{Kobayashi}},
  \bibinfo{author}{\bibfnamefont{O.}~\bibnamefont{Takeuchi}}, \bibnamefont{and}
  \bibinfo{author}{\bibfnamefont{H.}~\bibnamefont{Shigekawa}},
  \bibinfo{journal}{J. Am. Chem. Soc.} \textbf{\bibinfo{volume}{129}},
  \bibinfo{pages}{740} (\bibinfo{year}{2007}{\natexlab{a}}).

\bibitem[{\citenamefont{Temirov et~al.}(2006)\citenamefont{Temirov, Soubatch,
  Luican, and Tautz}}]{temirovn06}
\bibinfo{author}{\bibfnamefont{R.}~\bibnamefont{Temirov}},
  \bibinfo{author}{\bibfnamefont{S.}~\bibnamefont{Soubatch}},
  \bibinfo{author}{\bibfnamefont{A.}~\bibnamefont{Luican}}, \bibnamefont{and}
  \bibinfo{author}{\bibfnamefont{F.~S.} \bibnamefont{Tautz}},
  \bibinfo{journal}{Nature} \textbf{\bibinfo{volume}{444}},
  \bibinfo{pages}{350} (\bibinfo{year}{2006}).

\bibitem[{\citenamefont{Nicoara et~al.}(2006)\citenamefont{Nicoara, Roman,
  Gomez-Rodriguez, Martin-Gago, and Mendez}}]{Nicetal06}
\bibinfo{author}{\bibfnamefont{N.}~\bibnamefont{Nicoara}},
  \bibinfo{author}{\bibfnamefont{E.}~\bibnamefont{Roman}},
  \bibinfo{author}{\bibfnamefont{J.~M.} \bibnamefont{Gomez-Rodriguez}},
  \bibinfo{author}{\bibfnamefont{J.~A.} \bibnamefont{Martin-Gago}},
  \bibnamefont{and} \bibinfo{author}{\bibfnamefont{J.}~\bibnamefont{Mendez}},
  \bibinfo{journal}{Organic Electronics} \textbf{\bibinfo{volume}{7}},
  \bibinfo{pages}{287} (\bibinfo{year}{2006}).

\bibitem[{\citenamefont{Kevan}(1983)}]{kevanprb83}
\bibinfo{author}{\bibfnamefont{S.~D.} \bibnamefont{Kevan}},
  \bibinfo{journal}{Phys. Rev. B} \textbf{\bibinfo{volume}{28}},
  \bibinfo{pages}{2268} (\bibinfo{year}{1983}).

\bibitem[{\citenamefont{Euceda et~al.}(1983)\citenamefont{Euceda, Bylander, and
  Kleinman}}]{eucedaprb83}
\bibinfo{author}{\bibfnamefont{A.}~\bibnamefont{Euceda}},
  \bibinfo{author}{\bibfnamefont{D.~M.} \bibnamefont{Bylander}},
  \bibnamefont{and} \bibinfo{author}{\bibfnamefont{L.}~\bibnamefont{Kleinman}},
  \bibinfo{journal}{Phys. Rev. B} \textbf{\bibinfo{volume}{28}},
  \bibinfo{pages}{528} (\bibinfo{year}{1983}).

\bibitem[{\citenamefont{Baldacchini et~al.}(2003)\citenamefont{Baldacchini,
  Chiodo, Allegretti, Mariani, Betti, Monachesi, and {Del
  Sole}}}]{baldacchiniprb03}
\bibinfo{author}{\bibfnamefont{C.}~\bibnamefont{Baldacchini}},
  \bibinfo{author}{\bibfnamefont{L.}~\bibnamefont{Chiodo}},
  \bibinfo{author}{\bibfnamefont{F.}~\bibnamefont{Allegretti}},
  \bibinfo{author}{\bibfnamefont{C.}~\bibnamefont{Mariani}},
  \bibinfo{author}{\bibfnamefont{M.~G.} \bibnamefont{Betti}},
  \bibinfo{author}{\bibfnamefont{P.}~\bibnamefont{Monachesi}},
  \bibnamefont{and} \bibinfo{author}{\bibfnamefont{R.}~\bibnamefont{{Del
  Sole}}}, \bibinfo{journal}{Phys. Rev. B} \textbf{\bibinfo{volume}{68}},
  \bibinfo{pages}{195109} (\bibinfo{year}{2003}).

\bibitem[{\citenamefont{Sekiba et~al.}(2007)\citenamefont{Sekiba, Komori, and
  Cortona}}]{sekibaprb07}
\bibinfo{author}{\bibfnamefont{D.}~\bibnamefont{Sekiba}},
  \bibinfo{author}{\bibfnamefont{F.}~\bibnamefont{Komori}}, \bibnamefont{and}
  \bibinfo{author}{\bibfnamefont{P.}~\bibnamefont{Cortona}},
  \bibinfo{journal}{Phys. Rev. B} \textbf{\bibinfo{volume}{75}},
  \bibinfo{pages}{165410} (\bibinfo{year}{2007}).

\bibitem[{\citenamefont{Zhao et~al.}(1999)\citenamefont{Zhao, Gai, Zhao, Yang,
  and Sakurai}}]{zhaoss99}
\bibinfo{author}{\bibfnamefont{X.}~\bibnamefont{Zhao}},
  \bibinfo{author}{\bibfnamefont{Z.}~\bibnamefont{Gai}},
  \bibinfo{author}{\bibfnamefont{R.~G.} \bibnamefont{Zhao}},
  \bibinfo{author}{\bibfnamefont{W.~S.} \bibnamefont{Yang}}, \bibnamefont{and}
  \bibinfo{author}{\bibfnamefont{T.}~\bibnamefont{Sakurai}},
  \bibinfo{journal}{Surf. Sci.} \textbf{\bibinfo{volume}{424}},
  \bibinfo{pages}{L347} (\bibinfo{year}{1999}).

\bibitem[{\citenamefont{Zhao et~al.}(2001)\citenamefont{Zhao, Wang, Zhao, and
  Yang}}]{zhaomsec01}
\bibinfo{author}{\bibfnamefont{X.}~\bibnamefont{Zhao}},
  \bibinfo{author}{\bibfnamefont{H.}~\bibnamefont{Wang}},
  \bibinfo{author}{\bibfnamefont{R.~G.} \bibnamefont{Zhao}}, \bibnamefont{and}
  \bibinfo{author}{\bibfnamefont{W.~S.} \bibnamefont{Yang}},
  \bibinfo{journal}{Mater. Sci. Eng. C} \textbf{\bibinfo{volume}{16}},
  \bibinfo{pages}{41} (\bibinfo{year}{2001}).

\bibitem[{\citenamefont{Kang et~al.}(2003)\citenamefont{Kang, Toomes, Polcik,
  Kittel, Hoeft, Efstathiou, Woodruff, and Bradshaw}}]{kangjcp03}
\bibinfo{author}{\bibfnamefont{J.-H.} \bibnamefont{Kang}},
  \bibinfo{author}{\bibfnamefont{R.~L.} \bibnamefont{Toomes}},
  \bibinfo{author}{\bibfnamefont{M.}~\bibnamefont{Polcik}},
  \bibinfo{author}{\bibfnamefont{M.}~\bibnamefont{Kittel}},
  \bibinfo{author}{\bibfnamefont{J.-T.} \bibnamefont{Hoeft}},
  \bibinfo{author}{\bibfnamefont{V.}~\bibnamefont{Efstathiou}},
  \bibinfo{author}{\bibfnamefont{D.~P.} \bibnamefont{Woodruff}},
  \bibnamefont{and} \bibinfo{author}{\bibfnamefont{A.~M.}
  \bibnamefont{Bradshaw}}, \bibinfo{journal}{J. Chem. Phys.}
  \textbf{\bibinfo{volume}{118}}, \bibinfo{pages}{6059} (\bibinfo{year}{2003}).

\bibitem[{\citenamefont{Mae and Morikawa}(2004)}]{maess04}
\bibinfo{author}{\bibfnamefont{K.}~\bibnamefont{Mae}} \bibnamefont{and}
  \bibinfo{author}{\bibfnamefont{Y.}~\bibnamefont{Morikawa}},
  \bibinfo{journal}{Surf. Sci.} \textbf{\bibinfo{volume}{553}},
  \bibinfo{pages}{L63} (\bibinfo{year}{2004}).

\bibitem[{\citenamefont{Kanazawa
  et~al.}(2007{\natexlab{b}})\citenamefont{Kanazawa, Taninaka, Takeuchi, and
  Shigekawa}}]{kanazawaprl07}
\bibinfo{author}{\bibfnamefont{K.}~\bibnamefont{Kanazawa}},
  \bibinfo{author}{\bibfnamefont{A.}~\bibnamefont{Taninaka}},
  \bibinfo{author}{\bibfnamefont{O.}~\bibnamefont{Takeuchi}}, \bibnamefont{and}
  \bibinfo{author}{\bibfnamefont{H.}~\bibnamefont{Shigekawa}},
  \bibinfo{journal}{Phys. Rev. Lett.} \textbf{\bibinfo{volume}{99}},
  \bibinfo{eid}{216102} 
  (\bibinfo{year}{2007}.
  
\bibitem[{\citenamefont{Kresse and Furthm{\"u}ller}(1996)}]{kresseprb96}
\bibinfo{author}{\bibfnamefont{G.}~\bibnamefont{Kresse}} \bibnamefont{and}
  \bibinfo{author}{\bibfnamefont{J.}~\bibnamefont{Furthm{\"u}ller}},
  \bibinfo{journal}{Phys. Rev. B} \textbf{\bibinfo{volume}{54}},
  \bibinfo{pages}{11169} (\bibinfo{year}{1996}).

\bibitem[{\citenamefont{Kresse and Joubert}(1999)}]{kresseprb99}
\bibinfo{author}{\bibfnamefont{G.}~\bibnamefont{Kresse}} \bibnamefont{and}
  \bibinfo{author}{\bibfnamefont{D.}~\bibnamefont{Joubert}},
  \bibinfo{journal}{Phys. Rev. B} \textbf{\bibinfo{volume}{59}},
  \bibinfo{pages}{1758} (\bibinfo{year}{1999}).

\bibitem[{\citenamefont{Perdew and Wang}(1992)}]{perdewprb92}
\bibinfo{author}{\bibfnamefont{J.~P.} \bibnamefont{Perdew}} \bibnamefont{and}
  \bibinfo{author}{\bibfnamefont{Y.}~\bibnamefont{Wang}},
  \bibinfo{journal}{Phys. Rev. B} \textbf{\bibinfo{volume}{45}},
  \bibinfo{pages}{13244} (\bibinfo{year}{1992}).

\bibitem[{\citenamefont{Tersoff and Hamann}(1983)}]{tersoffprl83}
\bibinfo{author}{\bibfnamefont{J.}~\bibnamefont{Tersoff}} \bibnamefont{and}
  \bibinfo{author}{\bibfnamefont{D.~R.} \bibnamefont{Hamann}},
  \bibinfo{journal}{Phys. Rev. Lett.} \textbf{\bibinfo{volume}{50}},
  \bibinfo{pages}{1998} (\bibinfo{year}{1983}).

\bibitem[{\citenamefont{Barlow et~al.}(1998)\citenamefont{Barlow, Kitching,
  Haq, and Richardson}}]{barlowss98}
\bibinfo{author}{\bibfnamefont{S.~M.} \bibnamefont{Barlow}},
  \bibinfo{author}{\bibfnamefont{K.~J.} \bibnamefont{Kitching}},
  \bibinfo{author}{\bibfnamefont{S.}~\bibnamefont{Haq}}, \bibnamefont{and}
  \bibinfo{author}{\bibfnamefont{N.~V.} \bibnamefont{Richardson}},
  \bibinfo{journal}{Surf. Sci.} \textbf{\bibinfo{volume}{401}},
  \bibinfo{pages}{322} (\bibinfo{year}{1998}).

\end{thebibliography}
\end{document}